# The Potential of Singlet Fission Photon Multipliers as an Alternative to Silicon-based Tandem Solar Cells

*Moritz H. Futscher[1], Akshay Rao[2] and Bruno Ehrler[1]\**

1. Center for Nanophotonics, AMOLF, Science Park 104, 1098 XG Amsterdam, The Netherlands

2. Cavendish Laboratory, University of Cambridge, J.J. Thomson Avenue, Cambridge CB3 OHE, UK

**Corresponding Author**

* ehrler@amolf.nl

**Abstract**

Singlet fission, an exciton multiplication process in organic semiconductors which converts one singlet exciton into two triplet excitons is a promising way to reduce thermalization losses in conventional solar cells. One way to harvest triplet excitons is to transfer their energy into quantum dots, which then emit photons into an underlying solar cell. We simulate the performance potential of such a singlet fission photon multiplier combined with a silicon base cell and compare it to a silicon-based tandem solar cell. We calculate the influence of various loss-mechanisms on the performance potential under real-world operation conditions using a variety of silicon base cells with different efficiencies. We find that the photon multiplier is more stable against changes in the solar spectrum than two-terminal tandem solar cells. We furthermore find that, as the efficiency of the silicon solar cell increases, the efficiency of the photon multiplier increases at a higher rate than the tandem solar cell. For current record silicon solar cells, the photon multiplier has the potential to increase the efficiency by up to 4.2% absolute.

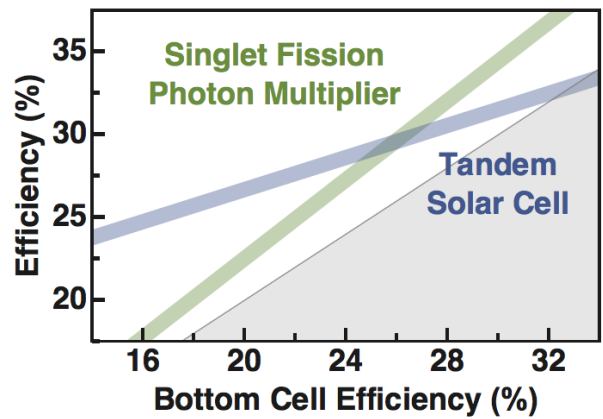

Crystalline silicon solar cells dominate the global solar-cell market and record efficiencies of 26.7% approach the Auger-recombination-constrained Shockley–Queisser limit.[1–4] For further improvement in the power-conversion efficiency new solutions beyond the silicon single-junction cell are needed.

Conventional solar cells lose a major part of incident sunlight energy via thermalization of excited charge carriers.[5] For a silicon solar cell with a band gap of 1.12 eV, thermalization accounts for a 39% power loss using the AM1.5G solar spectrum. The reduction of thermalization losses thus offers a great opportunity to achieve efficiencies above the Shockley-Queisser limit. Many strategies have been proposed to reduce thermalization losses of silicon solar cells, including tandem configurations and the modulation of the solar spectrum by down conversion.

In a tandem configuration with two sub-cells, a high-bandgap cell is placed on top of a low-bandgap cell.[6] Photons with a high energy are absorbed in the top cell and the transmitted light is absorbed in the bottom cell, reaching record efficiencies of 32.8% with III-V semiconductors as the top cell and silicon as the bottom cell in a four-terminal configuration.[7] Perovskites are a class of materials that promise cost-effective and efficient tandem solar cells in combination with silicon.[8–10] Perovskite/silicon tandem cells reach record efficiencies of 27.3%, exceeding the record efficiency of silicon solar cells.[11] However, tandem solar cells add extra costs and complexity to the fabrication process. They are furthermore sensitive to changes in solar spectrum and temperature during the course of a year, which reduces their efficiency under real-world conditions compared to laboratory conditions.[12,13]

While tandem solar cells are studied extensively, partially due to the recent boom in perovskite research, alternatives such as spectral modulation have received considerably less attention. Modulating the solar spectrum by either up- or down-conversion of photons,[14–18] single-junction solar cells can operate at an efficiency comparable to tandem solar cells.[15] A down-conversion device absorbs high-energy photons with at least twice the band gap energy, and emits twice as many photons with about half that energy. We call this device a "photon multiplier". Compared to tandem solar cells, a photon multiplier has the advantage that it could be integrated into existing solar cell technologies without the need for changes to the underlying solar cell, even as an upgrade (see Figure 1a).

Singlet fission, a spin-allowed exciton multiplication process in organic semiconductors which converts one singlet exciton into two triplet excitons,[19] is a suitable process for such a photon multiplier. Upon photoexcitation, organic semiconductors generate singlet excitons. If the energy of these singlet excitons $E(S_1)$ is close to twice the energy of the lowest-lying triplet exciton $E(T_1)$, i.e. $E(S_1) \approx 2\,E(T_1)$, singlet fission ($S_1 \rightarrow 2T_1$) can occur on sub-100 fs timescales.[20] Singlet fission has been observed with very high efficiency,[21] even for endothermic singlet fission, i.e. $E(S_1) < 2\,E(T_1)$. Endothermic singlet fission with barriers as high as 200 meV can still be highly efficient.[22] To form the photon multiplier, the triplet excitons can then transfer their energy into quantum dots (QD). Within the quantum dots the excitons recombine to emit photons,[23,24] whereby the exciton multiplication process becomes a photon multiplication process. Further details on singlet fission and the photon multiplier concept can be found in a recent review.[25]

The efficiency limit of singlet fission solar cells essentially matches the efficiency limit for a two-junction tandem solar cell.[26] However, these efficiency limits are calculated for ideal cells under standard test conditions, and both cell types have very different potential loss mechanisms, and a very different dependence on environmental conditions. Hence, here we simulate the potential performance of both types, but with realistic electrical and optical parameters, and simulate them under real-world environmental conditions using a variety of silicon bottom cells with different efficiencies. For the perovskite/silicon tandem solar cell, we opted for the monolithic two-terminal configuration, which is more attractive from an industrial point of view than the mechanically stacked four-terminal configuration.[27] Our simulations provide clear guidelines to optimize photon multiplier devices by including physical parameters such as the energy of the singlet exciton, the energy and the full width at half-maximum (FWHM) of the QD emission, losses due to absorption, transfer efficiency, and imperfect guiding of the emitted photons towards the bottom cell. We find that the photon multiplier is more stable against changing irradiation, spectral shape, and temperature than tandem solar cells, which are important factors in real world performance. We furthermore find that, as the efficiency of the silicon solar cell increases, the photon multiplier gains faster in efficiency than the tandem solar cell, and that for current record silicon solar cells, the photon multiplier has the potential to increase the efficiency by up to 4.2% absolute.

To simulate the performance of the singlet fission photon multiplier in combination with a silicon solar cell we model the modulation of the solar spectrum. The efficiency of the silicon solar cell is then calculated using our previously developed method and the modulated solar spectrum.[12] The silicon solar cell is modeled by including realistic solar-cell parameters such as Auger recombination, non-radiative recombination, and parasitic series and shunt resistance into detailed-balance calculations. To include parasitic absorption of the contacts we include the external quantum efficiency (EQE) in the model. This allows for simulating the performance of both the silicon solar cell and the photon multiplier for changing solar spectra and temperatures (see Supporting Information (SI) S1 for details).

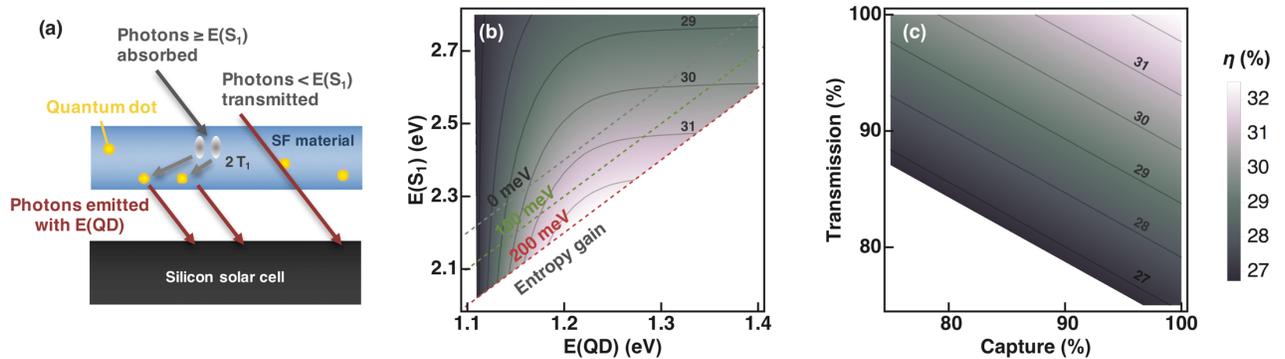

**Figure 1. (a)** Schematic illustration of the singlet fission photon multiplier device. **(b)** Efficiency of the singlet fission photon multiplier as a function of singlet exciton energy $E(S_1)$ and energy of quantum dot emission $E(QD)$ assuming no transmission and capture losses. **(c)** Efficiency of the singlet fission photon multiplier as a function of capture and transmission losses below $E(S_1)$. The capture parameter is defined in the text as $\eta_{SF} \times \eta_T \times \eta_{QD} \times \eta_C$. The calculations are performed at standard test conditions using a silicon base cell with an efficiency of 26.7%.

For a photon multiplier that absorbs all light above the energy of the singlet exciton $E(S_1)$ and where the QDs emit at the energy $E(QD) = ½ E(S_1)$ with a FWHM of 30 meV, the optimum energetics are $E(QD)$ 1.21 and $E(S_1)$ 2.42 eV. Including entropic gain, the optimal singlet exciton energy shifts to lower energies, while the energy for QD emission remains almost constant. At 200 meV entropic gain, the optimum would be at E(QD) = 1.22 eV and E(S$_1$) = 2.24 eV, where the efficiency of the record silicon solar cell would be enhanced from 26.7% to 32.5% (see Figure 1b).

To take transmission losses due to parasitic absorption and reflection by the photon multiplier into account, we assume that photons with an energy below E(S$_1$) are homogeneously absorbed or reflected before reaching the silicon solar cell. In addition, we consider losses during the photon multiplication process, which are collectively referred to as capture losses. These capture losses include the triplet exciton yield from singlet fission ($\eta_{SF}$), the efficiency of triplet excitons diffusing to, and transferring into the QDs ($\eta_T$), the photoluminescence quantum efficiency of the QDs ($\eta_{QD}$), and the fraction of photons emitted by the QDs towards the silicon solar cell ($\eta_C$). Ideally, the QDs are distributed evenly throughout the singlet fission material, spaced by the triplet diffusion length. Since the diffusion length for triplet excitons can be high,[28] only a small number of QDs need to be embedded in the singlet fission material. We estimate that quantum dots distributed in a micron-thick singlet fission layer, spaced by 50 nm, would absorb only 0.1% of the transmitted light, and therefore do not consider any reabsorption losses in our model. The efficiency of the singlet fission photon multiplier for different combinations of absorption and capture losses assuming a 30 meV FWHM for the QD emission and 200 meV entropic gain is shown in Figure 1c.

Table 1. Parameters and performance of the realistic and the optimistic singlet fission photomultiplier calculated at standard test conditions using a silicon base cell with an efficiency of 26.7%. The capture parameter is defined in the text as $\eta_{SF} \times \eta_T \times \eta_{QD} \times \eta_C$.

|  | Entropic gain (meV) | Transmission (%) | Capture (%) | FWHM (meV) | η (%) |
|---|---|---|---|---|---|
| **Realistic case** | 100 | 95 | 85 | 30 | 29.0 |
| **Optimistic case** | 200 | 97 | 95 | 30 | 31.3 |

In the following we compare two cases of photon multipliers. A relatively realistic case with an efficiency of 29.0% and an optimistic case with an efficiency of 31.3% using the record silicon base cell with an efficiency of 26.7% (see Table 1). The current-voltage characteristics of the modeled photon multiplier together with the modulated photon flux reaching the silicon solar cell filtered by its EQE are shown in Figure 2a. The effect of FWHM on the efficiency is relatively small, however, a wider emission spectrum does shift the ideal QD bandgap to slightly higher energies (Figure 2b, see also SI).

To compare the potential of the photon multiplier to tandem solar cells, we simulate a monolithic two-terminal perovskite/silicon tandem solar cell with all parameters based on the record perovskite solar cell with an efficiency of 22.7% and an area of 0.09 cm$^2$,[29] except that we change the bandgap to the ideal value of 1.68 eV, in order to current-match the perovskite top cell with the silicon bottom cell. The record perovskite cell features a shunt resistance of 5000 Ω cm$^2$, a series resistance of 0.32 Ω cm$^2$, and an electroluminescent emission efficiency of 0.15%. This leads to an efficiency of 20.9% for the perovskite solar cell and 32.7% for the tandem solar cell in combination with the record silicon solar cell with an efficiency of 26.7%

(see SI S2 for details). We note that we did not include any possible optical or electrical losses from the intermediate recombination layer required in practical tandem cells. The performance characteristics of the perovskite cell assumed here are hence optimistic and have not yet been achieved with this bandgap.

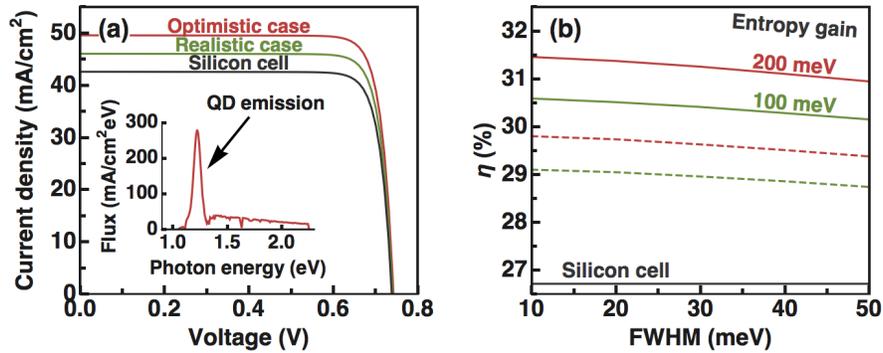

**Figure 2. (a)** Current-voltage characteristics of the modeled singlet fission photon multiplier on the silicon base cell with an efficiency of 26.7% for the optimistic case and the realistic case. The inset shows the modulated photon flux incident on the silicon base cell filtered by its external quantum efficiency. **(b)** Efficiency of the singlet fission photon multiplier as a function of full width at half-maximum (FWHM) for the quantum dot emission using a silicon base cell with an efficiency of 26.7%. The solid (dashed) lines assume 3% (5%) parasitic absorption losses below the singlet fission bandgap and capture losses of 5% (15%). The black solid line indicates the efficiency of the silicon solar cell.

To simulate real-world conditions we use solar spectra, irradiance, and temperatures measured in Utrecht, The Netherlands in 2015 at an interval of 30 min during daylight hours,[30] as described in previous work.[12]

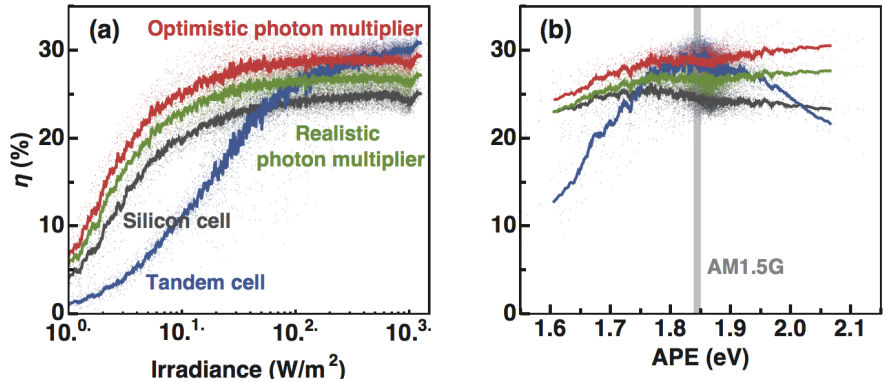

**Figure 3.** Efficiency of the two singlet fission photon multipliers, the tandem solar cell, and the silicon solar cell as a function of **(a)** irradiance and **(b)** average photon energy (APE) using solar spectra and temperatures measured in Utrecht, The Netherlands and the record silicon base cell with an efficiency of 26.7%.[30] The solid lines represent the moving average of the data. The APE is calculated for photons with an energy above the bandgap of silicon. The grey line indicates the APE of the AM1.5G standard solar spectrum.

Figure 3a shows that the efficiency of the tandem solar cell and the photon multipliers over the course of the year. Since the photon multiplier acts as a passive optical film modulating the incident solar spectrum, no electrical contact with the silicon solar cell is required. The photon multiplier thus shifts the silicon solar cell to higher efficiencies without considerably changing its dependence on the irradiance. The difference to the tandem cell is most prominent in the low-intensity region. The decrease in efficiency for tandem cells at low irradiance is due to the shunt resistance from the perovskite top cell that adds to the shunt resistance from the silicon cell alone. At high irradiances, the tandem solar cell overtakes the silicon solar cell and the two photon multipliers, as the increased shunt resistance at high current density is relatively less important.

Figure 3b shows that the efficiency of the silicon solar cell is only weakly affected by spectral changes, while the photon multiplier improves in efficiency with increasing average photon energy (APE). This is due to the modulation of the solar spectrum by the photon multiplier, which makes better use of the blue part of the incident solar spectrum. The tandem cell on the other hand is strongly affected by a shift of the APE away from standard test conditions, due to the current-matching constraint of the monolithic two-terminal configuration. In a monolithic two-terminal configuration, the generated current is limited by the cell producing the lower current. A change in the spectral irradiance distribution therefore leads to a discrepancy between the current generated in the two sub-cells, which reduces the efficiency of the tandem solar cell. We note that we only used values with an irradiance higher than 50 W/m$^2$ in Figure 3(b) to highlight the effect of APE on the efficiency, which would otherwise be skewed by the reduction in efficiency of the various cells at low intensity. Figure S5 in the SI shows the APE including all irradiances.

In addition to the record silicon solar cell, we simulate the performance of the two photon multipliers and the tandem solar cell using a variety of silicon base cells with (certified) efficiencies varying from 17.8% to 26.7% under standard test conditions. We also include two silicon solar cells with no non-radiative recombination, unity EQE, and with Auger recombination (29.9%) and without (30.6%). The bandgap of the perovskite solar cell, E(S$_1$), and E(QD) were optimized for each silicon solar cell (see SI S3 for details). Figure 4a shows the efficiencies of the tandem cell and the silicon cells with a photon multiplier, as a function of the silicon base cell efficiency under standard test conditions. The photon multiplier increases the current of the silicon solar cells by an almost constant percentage without changing the electrical properties. As a result, the absolute efficiency increase by the photon multiplier is almost constant for all silicon cells, and even increases slightly for more efficient silicon cells (slope 1.1 in Figure 4 for the realistic case and 1.3 for the optimistic case). That increase is due to the (on average) higher EQE of the efficient silicon cells close to the band-edge which allows for more efficient use of the photons emitted from the photon multiplier. In addition, the QD emission is slightly shifted towards the red for cells with a high EQE near the band edge, allowing for higher current gain.

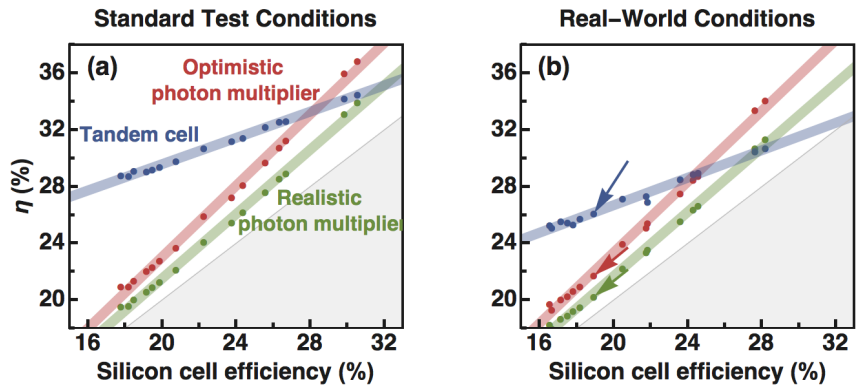

**Figure 4.** Efficiency of the two singlet fission photon multipliers and the tandem solar cell as a function of the silicon base cell efficiency under **(a)** standard test conditions and **(b)** real-world conditions averaged over the entire year and weighted with the incoming intensity, calculated using solar spectra and temperatures measured in Utrecht, The Netherlands.[30] The arrows indicate the change in efficiency from standard test conditions to real-world conditions.

In contrast, the tandem cell improves less upon the silicon cell efficiency for higher silicon efficiencies with a slope of 0.5 in Figure 4. This arises because the perovskite front cell shades part of the spectrum reaching the cell underneath, which leads to a larger loss for an efficient silicon base cell. The difference in efficiency between the tandem cell and the photon multiplier thus becomes lower the more efficient the silicon base cell and the photon multiplier becomes as efficient as the tandem cell at a silicon base cell efficiency of 28.2% for the optimistic and 32.0% for the realistic case under standard test conditions (see SI S3 for linear fit parameters of Figure 4).

Under realistic conditions, the decrease in efficiency compared to standard test conditions of the photon multiplier follows the efficiency drop of the silicon solar cell (see Figure 4b). This was already evident from Figure 3, because the dependence on irradiance and APE match the dependence of the silicon cell. In contrast, the efficiency of the tandem solar cell is strongly reduced under real-world conditions due to the sensitivity on changes in the solar spectrum. On average, the efficiencies of the silicon solar cells, the photon multipliers, and the tandem solar cells are reduced by 1.9, 2.0, and 3.7% absolute due to changes in the solar spectrum, irradiance, and temperature during the course of a year. The photon multiplier will then already be as efficient as the tandem solar cell at a silicon base-cell efficiency of 24.4 for the optimistic and 27.7% for the realistic case.

In conclusion, we simulate the performance potential of a singlet fission photon multiplier in comparison to a two-terminal tandem solar cell under real-world operation conditions. Compared to tandem solar cells, the photon multiplier has the advantage that it can be easily be integrated into existing solar cell technologies, without the need for electrical contacts with the underlying solar cell. Unlike monolithic two-terminal tandem cells, the photon multiplier does not require current-matching, making it more stable against changes in the solar spectrum.

To improve the efficiency of silicon solar cells by modulating the incident solar spectrum, however, some requirements must be met. The singlet fission material must have a high triplet exciton yield and a strong absorption. Furthermore, efficient energy transfer of the triplet excitons into the QDs is necessary, which must emit between 1.2 and 1.3 eV with a high photoluminescence quantum efficiency. A large proportion of the emitted photons must then be directed towards the underlying silicon solar cell. If this is achieved, we find that a photon multiplier can increase the efficiency of the record silicon solar cell by up to 4.2% absolute even at real-world environmental conditions. The purely optical method of modulating the incident solar spectrum with a singlet fission photon multiplier thus offers a promising way to reduce thermalization losses.

ASSOCIATED CONTENT

**Supporting Information**. Additional information is available on the singlet fission photon multiplier model, the perovskite/silicon tandem solar cell model, and the modeled silicon solar cells.

**Simulations**. Simulations of the solar cell efficiencies are performed using Mathematica. The code can be downloaded free of charge at https://github.com/HybridSolarCells.

AUTHOR INFORMATION

**Corresponding author**. ehrler@amolf.nl

**Group homepage**. https://amolf.nl/research-groups/hybrid-solar-cells

**Notes.** The authors declare no competing financial interest.

ACKNOWLEDGEMENT

The authors thank Wilfried van Sark and Atse Louwen from the Utrecht Photovoltaic Outdoor Test facility (UPOT) for providing spectral irradiance and temperature data and Erik Garnett for carefully reading and commenting on the manuscript. This work is part of the research program of the Netherlands Organization for Scientific Research (NWO).

SUPPLEMENTARY INFORMATION FOR

The Potential of Singlet Fission Photon Multipliers as an Alternative to Silicon-based Tandem Solar Cells

*Moritz H. Futscher[1], Akshay Rao[2] and Bruno Ehrler[1*]*

1. Center for Nanophotonics, AMOLF, Science Park 104, 1098 XG Amsterdam, The Netherlands

2. Cavendish Laboratory, University of Cambridge, J.J. Thomson Avenue, Cambridge CB3 OHE, UK

**Corresponding Author**

* ehrler@amolf.nl

## S1 SINGLET FISSION PHOTON MULTIPLIER MODEL

To model the modulation of the incident solar spectrum we assume that the singlet fission material absorbs everything above the energy of the singlet exciton $E(S_1)$ as

$$\Phi = \int_{E(S_1)}^{E_{max}} \Gamma(E)\, dE$$

where $\Gamma$ is the photon flux of the incident solar spectrum. Each absorbed photon generates one singlet exciton with an energy $E(S_1)$ which converts into two triplet excitons with an energy $E(T_1)$. The energy of the triplet excitons is then transferred into quantum dots (QD) which emit photons with an energy of $E(QD)$ and a full width at half-maximum (FWHM) of $\sigma_{QD}$ as

$$\Gamma_{\text{gain}}(E) = \frac{2\,\Phi\,\gamma}{\sqrt{2\pi}\,\sigma_{QD}}\, e^{-\frac{(E-E(QD))^2}{2\,\sigma_{QD}^2}}$$

where $\gamma$ is the capture parameter including the efficiency of triplet exciton yield form singlet fission, the efficiency of triplet excitons transferring their energy to QDs, the photoluminescence quantum efficiency of the QDs, and the efficiency of photons emitted by the QDs reaching the silicon solar cell, i.e. $\gamma = \eta_{SF}\,\eta_T\,\eta_{QD}\,\eta_C$. In the ideal case, the capture parameter is unity. The generated photocurrent density in the bottom cell can then be calculated as

$$J_G = q \int_{E_{min}}^{E(S_1)} [\,\zeta\,\Gamma(E) + \Gamma_{\text{gain}}(E)\,]\, \text{EQE}(E)\, dE$$

where $\zeta$ is the fraction of light transmitted through the photon multiplier and EQE is the external quantum efficiency of the bottom cell to take additional optical losses such as parasitic absorption in the bottom cell contacts into account.

The current-voltage characteristic of a silicon solar cell can then be calculated following previous work as[1]

$$J(V) = J_G - J_R \left( e^{\frac{q(V+JR_S)}{k_B T}} - 1 \right) - J_{NR} \left( e^{\frac{q(V+JR_S)}{k_B T}} - 1 \right) - J_A \left( e^{\frac{3q(V+JR_S)}{2k_B T}} - 1 \right)$$

$$- \frac{V + JR_s}{R_{sh}}$$

where $V$ the applied voltage, $k_B$ the Boltzmann constant, $T$ the temperature of the cell, $R_s$ is the series resistance, and $R_{sh}$ the shunt resistance. The second term of $J(V)$ corresponds to the radiative recombination current density, the third term to the non-radiative recombination current density, the fourth term to the Auger recombination current density, and the last term is due to parasitic resistances in the cell. The electroluminescent emission efficiency ($\eta_{EL}$) is calculated as[2]

$$J(V) = J_G - \frac{J_R}{\eta_{EL}} \left( e^{\frac{q(V+JR_S)}{k_B T}} - 1 \right) - \frac{V + JR_s}{R_{sh}}$$

where $\eta_{EL}$ is calculated as

$$\eta_{EL} = \frac{J_R}{J_R + J_{NR} + J_A \left( 1 + e^{\frac{3q(V+JR_S)}{2k_B T}} \right)^{-1}}.$$

The current-voltage characteristics of a silicon solar cell can then be fitted by including the EQE and the silicon thickness $L$, and adjusting the amount of non-radiative recombination $J_{NR}$, $R_s$, and $R_{sh}$. The current-voltage characteristics of the modeled record silicon solar cells together with the fitting parameters and its EQE is shown in Figure S1.

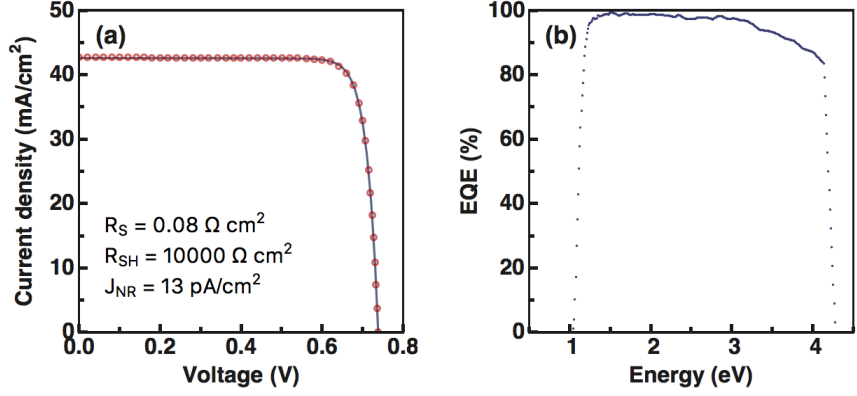

**Figure S1. (a)** Current–voltage characteristics of the record efficiency silicon solar cells with an area of 79 cm$^2$, a $V_{OC}$ of 0.738 V, a $J_{SC}$ of 42.65 mA/cm$^2$, a fill factor of 84.9%, an $\eta_{EL}$ of 0.4%, and an efficiency of 26.7%.[3] The circles correspond to the measured data of the record efficiency silicon solar cell and the solid line corresponds to the fitted current-voltage characteristics. **(b)** External quantum efficiency of the record silicon solar cell used to account for optical losses.

## S2 TANDEM SOLAR CELL MODEL

To model a perovskite solar cell with an ideal bandgap for a tandem solar cell in combination with silicon, we first fit the current-voltage characteristic of the current record perovskite solar with a bandgap of 1.49 eV, an area of 0.09 cm$^2$, and an efficiency of 22.7%.[4] We calculate the current-voltage characteristic of a perovskite solar cell following previous work as[1]

$$J(V) = J_G - J_R\left(e^{\frac{q(V+JR_S)}{k_BT}} - 1\right) - J_{NR}\left(e^{\frac{q(V+JR_S)}{2k_BT}} - 1\right) - \frac{V+JR_S}{R_{sh}}.$$

The electroluminescent emission efficiency for the perovskite cell is calculated as

$$\eta_{EL} = \frac{J_R}{J_R + J_{NR}(V)\left(1 + e^{\frac{q(V+JR_S)}{2k_BT}}\right)^{-1}}.$$

The current-voltage characteristics of a perovskite solar cell can then be modeled by including the EQE and adjusting the amount of non-radiative recombination $J_{NR}$, $R_s$, and

$R_{sh}$. The current-voltage characteristics of the modeled record perovskite solar cells together with the fitting parameters and its EQE is shown in Figure S2.

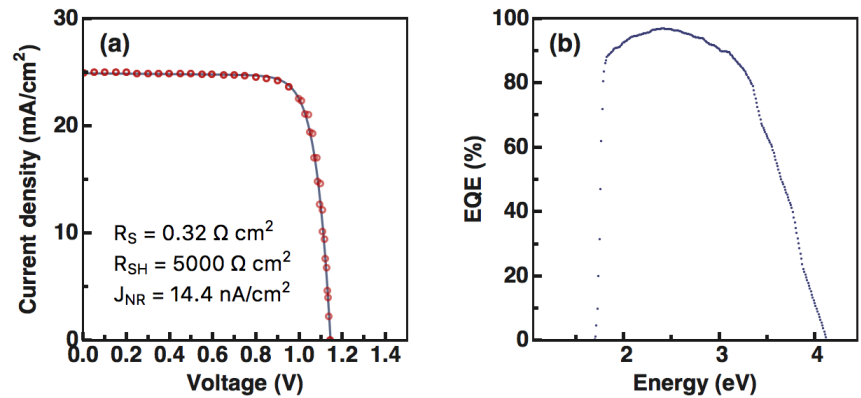

**Figure S2. (a)** Current–voltage characteristics of the record efficiency perovskite solar cells with a bandgap of 1.49 eV, an area of 0.09 cm$^2$, a $V_{OC}$ of 1.144 V, a $J_{SC}$ of 24.91 mA/cm$^2$, a fill factor of 79.6%, an $\eta_{EL}$ of 0.15% at maximum power point (MPP), and an efficiency of 22.7%.[4] The circles correspond to the measured data of the record efficiency perovskite solar cell and the solid line corresponds to the modeled current-voltage characteristics. **(b)** External quantum efficiency of the record perovskite solar cell used to account for optical losses.

To optimize for maximum perovskite/silicon tandem solar cell efficiency, we change the bandgap of the perovskite solar cell artificially by scaling the EQE of the record perovskite solar cell in energy until current matching between the perovskite top cell and the silicon bottom cell is achieved while keeping the parasitic resistances and the electroluminescent emission efficiency constant. The modeled perovskite solar cell then has a bandgap of 1.68 eV and an efficiency of 20.9%. The current-voltage characteristics of the modeled perovskite solar cell with an ideal bandgap together with the fitting parameters and its EQE is shown in Figure S3.

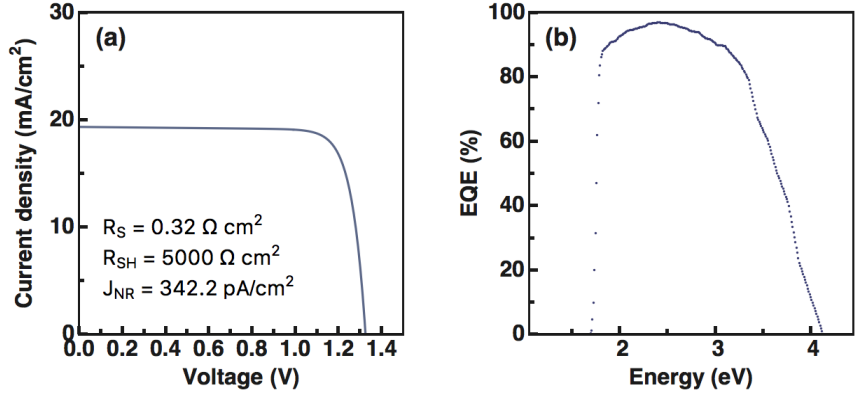

**Figure S3.** (a) Current–voltage characteristics and (b) external quantum efficiency of the perovskite solar cells with a bandgap of 1.68 eV, a $V_{OC}$ of 1.325 V, a $J_{SC}$ of 19.36 mA/cm$^2$, a fill factor of 81.5%, an $\eta_{EL}$ of 0.15% at MPP, and an efficiency of 20.9% used to model the perovskite/silicon tandem solar cell.

To model the perovskite/silicon tandem solar cell we apply modifications to the detailed-balance limit following previous work.[5] The power for the monolithic two-terminal tandem solar cell is calculated as

$$P_{out} = (V_{Perovskite} + V_{Silicon})J$$

The conversion efficiency for perovskite/silicon tandem solar cells is then given by

$$\eta = Max\left(\frac{P_{out}(V)}{P_{sun}}\right)$$

where $P_{sun}$ is the power of the incident solar spectrum. Optical losses are included by fitting a Gaussian distribution to the onset of the EQE spectra of the perovskite cell. 10% of the light with an energy below the Gaussian distribution is assumed to be uniformly absorbed by parasitic absorption in the perovskite contacts. The current-voltage characteristics of the modeled perovskite/silicon tandem solar cells together the EQE of the optimistic perovskite and the record silicon subcell is shown in Figure S4.

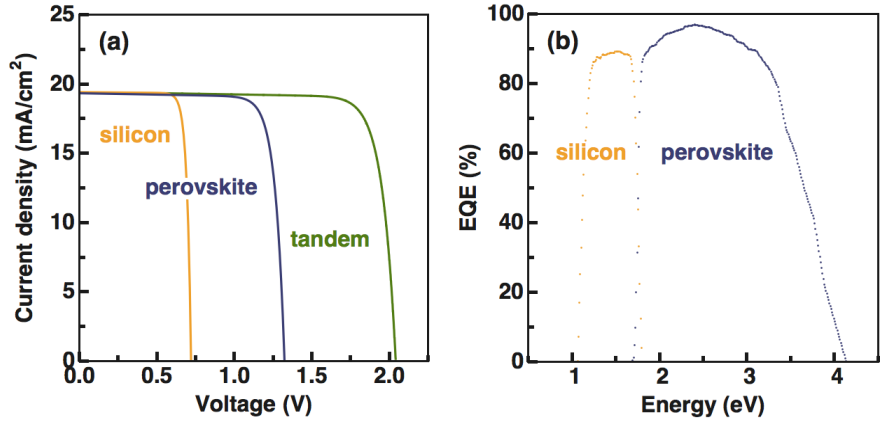

**Figure S4. (a)** Current-voltage characteristics and **(b)** external quantum efficiency of the modeled monolithic two-terminal perovskite/silicon tandem solar cell with an efficiency of 32.7% together with its sub-cells.

## S3 MODELED SILICON SOLAR CELLS

To simulate the performance of a photon multiplier and a tandem solar cell as a function of the silicon solar cell efficiency, we fit our model to a variety of certified silicon solar cells. The parameters of the modeled silicon solar cells are shown in Table 1. Cell 1 to cell 13 are modeled after previous record silicon solar cells including monocrystalline and polycrystalline silicon solar cell, with cell 13 being the current record silicon solar cell. Cell 14 and 15 are hypothetical, idealized cells. Cell 14 and cell 15 both assume an ideal EQE of 100% above the silicon band gap, and cell 15 additionally assumes no Auger recombination.

To optimize for maximum efficiency of the photon multiplier and the tandem solar cell, we optimize the bandgap of the perovskite solar cell, the absorption edge of the singlet fission material $E(S_1)$, and $E(QD)$ for each modeled silicon solar cell. The parameters for the optimistic and the realistic photon multiplier for each silicon solar cell are shown in Table 2 and Table 3, respectively. The parameters for the optimized perovskite solar cells for each silicon solar cell are shown in Table 4. The effect of FWHM on the efficiency is shown in Table 5 and Table 6. Table 7 and Table 8 furthermore show the fitting parameters to Figure 4 in the main text.

**Table 1.** Summary of modeled silicon solar cells calculated at standard test conditions (AM1.5G, 25 °C, 1 kWh/m$^2$). Thicknesses of active material L marked with an asterisk are estimates.

| Cell | Source | $V_{OC}$ (V) | $J_{SC}$ (mA/cm$^2$) | FF (%) | η (%) | $R_S$ (Ωcm$^2$) | $R_{SH}$ (Ωcm$^2$) | $J_{NR}$ (pA/cm$^2$) | $η_{EL}$ (‰) | L (μm) |
|---|---|---|---|---|---|---|---|---|---|---|
| 1 | [6] | 0.648 | 36.29 | 75.7 | 17.8 | 1.48 | 1800 | 400 | 0.13 | 170 |
| 2 | [7] | 0.638 | 37.19 | 76.7 | 18.2 | 1.15 | 1300 | 591 | 0.09 | 180* |
| 3 | [8] | 0.650 | 37.38 | 76.2 | 18.5 | 1.35 | 1800 | 386 | 0.13 | 180 |
| 4 | [9] | 0.649 | 37.50 | 78.9 | 19.2 | 0.82 | 1800 | 400 | 0.13 | 190 |
| 5 | [10] | 0.692 | 36.41 | 77.4 | 19.5 | 1.28 | 1200 | 72 | 0.71 | 180* |
| 6 | [11] | 0.657 | 38.10 | 79.5 | 19.9 | 0.73 | 2000 | 297 | 0.17 | 180* |
| 7 | [12] | 0.663 | 39.03 | 80.3 | 20.8 | 0.59 | 2000 | 241 | 0.21 | 190 |
| 8 | [13] | 0.674 | 41.07 | 80.5 | 22.3 | 0.60 | 3000 | 166 | 0.31 | 195 |
| 9 | [10] | 0.742 | 39.31 | 81.6 | 23.8 | 0.67 | 2000 | 10 | 4.98 | 150 |
| 10 | [11] | 0.736 | 41.39 | 80.1 | 24.4 | 0.75 | 800 | 13 | 3.84 | 200 |
| 11 | [14] | 0.740 | 41.81 | 82.7 | 25.6 | 0.47 | 4000 | 12 | 4.4 | 150 |
| 12 | [11] | 0.744 | 42.26 | 83.8 | 26.4 | 0.31 | 6000 | 10 | 5.3 | 200 |
| 13 | [15] | 0.738 | 42.65 | 84.9 | 26.7 | 0.08 | 10000 | 13 | 4.0 | 200 |
| 14 | - | 0.776 | 43.61 | 88.3 | 29.9 | 0.01 | 10000 | 1 | 69.9 | 200 |
| 15 | - | 0.814 | 43.61 | 86.1 | 30.6 | 0.01 | 1000 | 1 | 69.9 | 200 |

**Table 2.** Summary of parameters for the optimistic singlet fission photon multiplier and the resulting solar cell for each silicon bottom cell calculated at standard test conditions. The optimistic case assumes 200 meV entropy gain, 3% parasitic absorption losses below the singlet fission bandgap, capture losses of 5%, and a FWHM for the QD emission of 30 meV.

| Cell | $E(S_1)$ (eV) | $E(QD)$ (eV) | $V_{OC}$ (V) | $J_{SC}$ (mA/cm$^2$) | FF (%) | η (%) |
|---|---|---|---|---|---|---|
| 1 | 2.20 | 1.20 | 0.652 | 43.15 | 74.5 | 21.0 |
| 2 | 2.26 | 1.23 | 0.642 | 42.91 | 76.0 | 20.9 |
| 3 | 2.28 | 1.24 | 0.654 | 43.51 | 75.3 | 21.4 |
| 4 | 2.32 | 1.26 | 0.652 | 43.14 | 78.4 | 22.1 |
| 5 | 2.3 | 1.25 | 0.695 | 41.90 | 76.8 | 22.4 |
| 6 | 2.32 | 1.26 | 0.660 | 43.60 | 79.10 | 22.8 |
| 7 | 2.32 | 1.26 | 0.666 | 44.50 | 80.0 | 23.7 |
| 8 | 2.26 | 1.23 | 0.678 | 47.76 | 80.1 | 25.9 |
| 9 | 2.30 | 1.25 | 0.754 | 45.05 | 81.3 | 27.3 |
| 10 | 2.26 | 1.23 | 0.740 | 47.68 | 79.8 | 28.1 |
| 11 | 2.28 | 1.24 | 0.744 | 48.48 | 82.5 | 29.7 |
| 12 | 2.26 | 1.23 | 0.748 | 49.22 | 83.7 | 30.8 |
| 13 | 2.24 | 1.22 | 0.742 | 49.63 | 85.0 | 31.3 |
| 14 | 2.18 | 1.19 | 0.780 | 52.29 | 88.4 | 36.0 |
| 15 | 2.18 | 1.19 | 0.819 | 52.29 | 86.2 | 36.9 |

**Table 3.** Summary of parameters for the realistic singlet fission photon multiplier and the resulting solar cell for each silicon bottom cell calculated at standard test conditions. The realistic case assumes 100 meV entropy gain, 5% parasitic absorption losses below the singlet fission bandgap, capture losses of 15%, and a FWHM for the QD emission of 30 meV.

| Cell | $E(S_1)$ (eV) | $E(QD)$ (eV) | $V_{OC}$ (V) | $J_{SC}$ (mA/cm$^2$) | FF (%) | η (%) |
|---|---|---|---|---|---|---|
| 1 | 2.30 | 1.20 | 0.651 | 40.04 | 75.1 | 19.5 |
| 2 | 2.36 | 1.23 | 0.640 | 40.06 | 76.4 | 19.6 |
| 3 | 2.38 | 1.24 | 0.652 | 40.61 | 75.7 | 20.0 |
| 4 | 2.42 | 1.26 | 0.651 | 40.29 | 78.6 | 20.6 |
| 5 | 2.40 | 1.25 | 0.694 | 39.14 | 77.1 | 20.9 |
| 6 | 2.42 | 1.26 | 0.659 | 40.72 | 79.3 | 21.3 |
| 7 | 2.42 | 1.26 | 0.664 | 41.60 | 80.2 | 22.1 |
| 8 | 2.36 | 1.23 | 0.676 | 44.43 | 80.3 | 24.1 |
| 9 | 2.42 | 1.26 | 0.743 | 42.07 | 81.5 | 25.5 |
| 10 | 2.36 | 1.23 | 0.738 | 44.42 | 80.0 | 26.2 |
| 11 | 2.38 | 1.24 | 0.742 | 45.11 | 82.6 | 27.6 |
| 12 | 2.36 | 1.23 | 0.746 | 45.73 | 83.8 | 28.6 |
| 13 | 2.34 | 1.22 | 0.740 | 46.10 | 84.9 | 29.0 |
| 14 | 2.28 | 1.19 | 0.778 | 48.24 | 88.3 | 33.1 |
| 15 | 2.28 | 1.19 | 0.817 | 48.24 | 86.2 | 33.9 |

**Table 4.** Summary of the parameters for the modeled perovskite solar cell for each silicon bottom cell calculated at standard test conditions, together with the tandem solar cell efficiency. All modeled perovskite solar cells have a shunt resistance of 5000 $\Omega$ cm$^2$, a series resistance of 0.32 $\Omega$ cm$^2$, and an $\eta_{EL}$ of 0.15% at MPP.

| Cell | Bandgap (eV) | $J_{NR}$ (pA/cm$^2$) | $V_{OC}$ (V) | $J_{SC}$ (mA/cm$^2$) | FF (%) | $\eta$ (%) | $\eta_{Tandem}$ (%) |
|---|---|---|---|---|---|---|---|
| 1 | 1.74 | 104 | 1.382 | 17.76 | 81.9 | 20.1 | 28.8 |
| 2 | 1.74 | 104 | 1.382 | 17.76 | 81.9 | 20.1 | 28.8 |
| 3 | 1.73 | 125 | 1.373 | 18.02 | 81.9 | 20.3 | 29.1 |
| 4 | 1.74 | 104 | 1.382 | 17.76 | 81.9 | 20.1 | 29.1 |
| 5 | 1.75 | 87 | 1.390 | 17.50 | 82.0 | 19.9 | 29.2 |
| 6 | 1.73 | 125 | 1.373 | 18.02 | 81.9 | 20.3 | 29.4 |
| 7 | 1.72 | 153 | 1.363 | 18.28 | 81.8 | 20.4 | 29.8 |
| 8 | 1.70 | 233 | 1.343 | 18.81 | 81.6 | 20.6 | 30.7 |
| 9 | 1.72 | 153 | 1.363 | 18.28 | 81.8 | 20.4 | 31.2 |
| 10 | 1.70 | 233 | 1.343 | 18.81 | 81.6 | 20.6 | 31.5 |
| 11 | 1.69 | 286 | 1.333 | 19.08 | 81.6 | 20.7 | 32.2 |
| 12 | 1.68 | 342 | 1.325 | 19.36 | 81.5 | 20.9 | 32.6 |
| 13 | 1.68 | 342 | 1.325 | 19.36 | 81.5 | 20.9 | 32.7 |
| 14 | 1.66 | 500 | 1.307 | 19.90 | 81.3 | 21.1 | 34.3 |
| 15 | 1.66 | 500 | 1.307 | 19.90 | 81.3 | 21.1 | 34.5 |

**Table 5.** Summary of parameters for the optimistic singlet fission photon multiplier and the resulting solar cell for different full width at half-maxima (FWHM) calculated at standard test conditions. The optimistic case assumes 200 meV entropy gain, 3% parasitic absorption losses below the singlet fission bandgap, capture losses of 5%, and a FWHM for the QD emission of 30 meV.

| FWHM (meV) | $E(S_1)$ (eV) | $E(QD)$ (eV) | $V_{OC}$ (V) | $J_{SC}$ (mA/cm$^2$) | FF (%) | η (%) |
|---|---|---|---|---|---|---|
| 10 | 2.22 | 1.21 | 0.742 | 49.94 | 85.0 | 31.5 |
| 20 | 1.24 | 1.22 | 0.742 | 49.81 | 85.0 | 31.4 |
| 30 | 1.24 | 1.22 | 0.742 | 49.63 | 85.0 | 31.3 |
| 40 | 2.26 | 1.23 | 0.742 | 49.40 | 85.0 | 31.1 |
| 50 | 2.26 | 1.23 | 0.741 | 49.16 | 85.0 | 30.9 |

**Table 6.** Summary of parameters for the realistic singlet fission photon multiplier and the resulting solar cell for different full width at half-maxima (FWHM) calculated at standard test conditions. The realistic case assumes 100 meV entropy gain, 5% parasitic absorption losses below the singlet fission bandgap, capture losses of 15%, and a FWHM for the QD emission of 30 meV.

| FWHM (meV) | $E(S_1)$ (eV) | $E(QD)$ (eV) | $V_{OC}$ (V) | $J_{SC}$ (mA/cm$^2$) | FF (%) | η (%) |
|---|---|---|---|---|---|---|
| 10 | 2.32 | 1.21 | 0.740 | 46.31 | 84.9 | 29.0 |
| 20 | 2.34 | 1.22 | 0.740 | 46.23 | 84.9 | 29.0 |
| 30 | 2.34 | 1.22 | 0.740 | 46.10 | 84.9 | 29.0 |
| 40 | 2.36 | 1.23 | 0.740 | 45.94 | 84.9 | 28.9 |
| 50 | 2.36 | 1.23 | 0.740 | 45.76 | 84.9 | 28.7 |

**Table 7.** Linear fitting parameters for the efficiency of the two singlet fission photon multipliers and the tandem solar cell as a function of the silicon solar cell efficiency under standard test conditions.

|  | Offset | Slope |
|---|---|---|
| **Realistic photon multiplier** | -1.20 | 1.14 |
| **Optimistic photon multiplier** | -2.12 | 1.26 |
| **Tandem solar cell** | 20.32 | 0.46 |

**Table 8.** Linear fitting parameters for the efficiency of the two singlet fission photon multipliers and the tandem solar cell as a function of the silicon solar cell efficiency under real-world conditions calculated using solar spectra and temperatures measured in Utrecht, The Netherlands.[16]

|  | Offset | Slope |
|---|---|---|
| **Realistic photon multiplier** | -0.96 | 1.13 |
| **Optimistic photon multiplier** | -1.62 | 1.25 |
| **Tandem solar cell** | 17.00 | 0.48 |

## S4 THE EFFECT OF APE ON THE EFFICIENCY

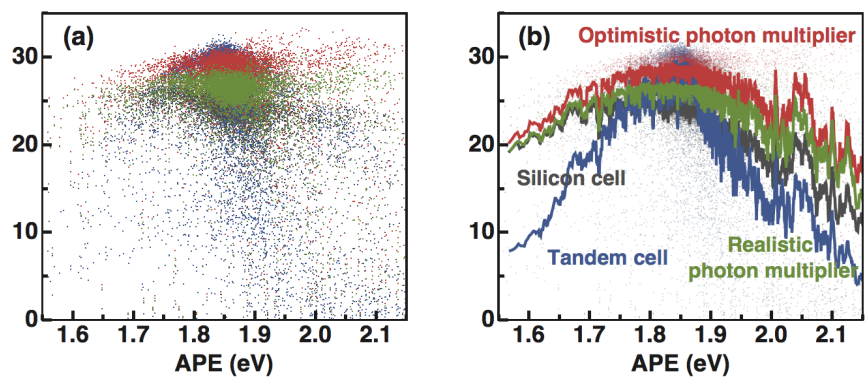

**Figure S5.** Efficiency of the two singlet fission photon multipliers, the tandem solar cell, and the silicon base cell as a function of average photon energy (APE) using solar spectra and temperatures measured in Utrecht, The Netherlands[16] and the record silicon base cell with an efficiency of 27.6%. The APE is calculated for photons with an energy above the bandgap of silicon. The solid line in **(b)** represents a moving average of the data shown in **(a)**.